# Addressing Hate Speech with Data Science: An Overview from Computer Science Perspective

*Ivan Srba, Gabriele Lenzini, Matus Pikuliak, Samuel Pecar*

## Keywords

## Abstract

From a computer science perspective, addressing on-line hate speech is a challenging task that is attracting the attention of both industry (mainly social media platform owners) and academia. In this chapter, we provide an overview of state-of-the-art data-science approaches – how they define hate speech, which tasks they solve to mitigate the phenomenon, and how they address these tasks. We limit our investigation mostly to (semi-)automatic detection of hate speech, which is the task that the majority of existing computer science works focus on. Finally, we summarize the challenges and the open problems in the current data-science research and the future directions in this field. Our aim is to prepare an easily understandable report, capable to promote the multidisciplinary character of hate speech research. Researchers from other domains (e.g., psychology and sociology) can thus take advantage of the knowledge achieved in the computer science domain but also contribute back and help improve how computer science is addressing that urgent and socially relevant issue which is the prevalence of hate speech in social media.

## 1 Introduction

The debate about the role of the social media (e.g., social networks, discussion forums, blogs or news sites) in modern societies is, more and more often, infused with criticism because of the prevalence of various types of antisocial behaviour or general misbehaviour. Such types of undesired behaviour hinder the social media's initial purpose: to help people stay in contact and share experiences and knowledge. Social media have, in fact, become also means to discriminate, polarize, offend, and denigrate ethnic, cultural, or national minorities (Mondal, Silva, & Benevenuto, 2017). *Hate speech* is the term used to indicate this incitement to violence and hatred against a specific category of persons, although, when referred to social media, we should more precisely be saying *online hate speech*. Among the various acts of misbehaviour, hate speech has been recognized as being one of the most pervasive and serious (Mathew, Dutt, Goyal, & Mukherjee, 2019).

In the EU, hate speech is considered an instance of hate crime, hence a criminal offence and an expression of discrimination (European Union Agency for Fundamental Rights, 2016). The substantial negative consequences of hate speech have prompted a significant effort to find appropriate barriers to its spreading that also do not hamper freedom of expression Without entering into detail, there is an evident tension between freedom of speech, from the one side, and protecting people from hate, from the other side. Even the instruments to preserve one's right are debated: anonymity can protect individuals from being unlawfully surveyed and thus to become targets of hate speech, but the same instrument may worsen the phenomenon by allowing perpetrators to hide and escape public scrutiny (Mondal et al., 2017).

The effort to mitigate hate speech is therefore motivated not only by the practical reasons of user retainment (who may decide to leave a hateful social media), but also by legislation. In many countries, site owners are compelled to react promptly and remove instances of hate speech as soon as they are recognized. To be precise, the EU has addressed the matter of hate speech in several occasions, e.g., in the Int. Convention on the Elimination of Racial Discrimination (1965), in the Int. Covenant of Civil and Political Rights (1996), and more recently in the recommendations elaborated by the European Commission against Racism and Intolerance (ECRI). In 2016, with its European Code of Conduct on Countering Illegal Hate Speech, the European Commission agreed with Facebook, YouTube, Twitter and Microsoft to counter the spread of hate speech on-line and to remove or disable access to hate speech content within 24 hours. In 2018 and 2019, Instagram, Google+, Snapchat, Dailymotion and Jeuxvideo.com joined the Code of Conduct.

Social media platforms employ various policies (e.g., terms of service or community guidelines) to define hate speech and to introduce actions to eliminate, or at least unlink, such content (Pater, Kim, Mynatt, & Fiesler, 2016). For instance, Facebook (other platforms adopt a similar phrasing in their usage policies), one of the subscribers to the EU Code of Conduct, states that:

> "We do not allow hate speech [...] because it creates an environment of intimidation and exclusion, and in some cases, may promote real-world violence. We define hate speech as a direct attack on people based on what we call protected characteristics – race, ethnicity, national origin, religious affiliation, sexual orientation, caste, sex, gender, gender identity and serious disease or disability. We also provide some protections for immigration status. We define "attack" as violent or dehumanising speech, statements of inferiority, or calls for exclusion or segregation." (Facebook, Community Standards)

To recognize hate speech, some social media platforms employ humans i.e., moderators and/or crowdsourcing. Deciding whether a content (usually a piece of text, such as a discussion post, a comment or a post at social networking site) is hateful, requires extensive experience, cultural awareness, and knowledge of human affairs (Waseem, 2016). The human-driven approach comes with strong limitations, though. Even expert humans are not exempted from cognitive and cultural biases and, thus, they need guidelines and exact definitions helping them distinguish hateful from benign content. Besides, let people peruse speeches do not scale up and can hardly cope with the gargantuan amount of daily user-generated content.

As a result, there is a strong demand to replace, at least in the most obvious cases, or to support, in the most general, the human-based detection with an automatic or semi-automatic detection; and, specifically, there is an interest in *data-science solutions*. Boosted by innovative Artificial Intelligence (AI) methods, data science promises to detect the most probable comments containing hate speech as early as possible after they are published.

Despite a high demand, there is so far only a low number of research studies and approaches in this field. In this chapter, we provide an overview of the state-of-the-art data-science approaches focused on hate speech. In addition, we discuss challenges and open problems in an understandable way for researchers also from other domains. To the best of our knowledge, the most recent comprehensive survey on hate speech from a computer science (CS) perspective is that by (Fortuna & Nunes, 2018). Published in 2018, it however misses the very recent progress in this area, whereas we take into consideration also the recent rapid advances in artificial neural networks, deep learning and Natural Language Processing (NLP).

## 2 Hate speech from a computer science perspective

### 2.1 Definition and Categorization

In spite of the attention from industry and academia on the problem of hate speech, there is no agreement on what is hate speech precisely. Both CS research (Fortuna & Nunes, 2018; Waseem, Davidson, Warmsley, & Weber, 2017) and social media platforms (Pater et al., 2016, Fortuna & Nunes, 2018) propose several definitions without agreeing on one. Here, we refer to the definition by (Fortuna & Nunes, 2018). They analysed a number of hate speech definitions from a wide spectrum of resources (including research papers, legislation and social media platforms) and proposed the following one (p. 5):

*Definition 1 (Hate Speech).* Hate speech is language that attacks or diminishes, that incites violence or hate against groups, based on specific characteristics such as physical appearance, religion, descent, national or ethnic origin, sexual orientation, gender identity or other, and it can occur with different linguistic styles, even in subtle forms or when humour is used.

The CS literature is even less consistent when situating hate speech into the context of other related and superior concepts and distinguishing the differences between them (Waseem et al., 2017). At first, hate speech is considered as a special type of *abusive language* (Fortuna & Nunes, 2018; Waseem et al., 2017). Some works (Davidson, Wamsley, Macy, & Weber, 2017; Zampieri et al., 2019) use a term *offensive language* interchangeably. Besides hate speech, other types of abusive/offensive language include cyberbullying, trolling or toxic language. Hate speech is also associated with broader concepts of *online harassment* (Pater et al., 2016; Wachs, Wright, & Vazsonyi, 2019) and *antisocial behaviour* (Chandra, Khatri, & Som, 2017; Švec, Pikuliak, Šimko, & Bieliková, 2018).

Since the above-mentioned different types of abusive language overlap significantly, many existing works do not distinguish between them. (Waseem et al., 2017) chose not to define terms but propose an alternative typology that may be more effective at identifying particular types of abuse. This typology differentiates abusive language on the basis of (a) whether the language is directed towards a specific individual or entity, or towards a more generalized group; and (b) whether the abusive content is explicit or implicit.

Besides, a definition of hate speech is also influenced by another practical aspect: namely, how existing datasets are labelled. If researchers work with some datasets coming from social media platforms, such datasets may be already annotated by moderators or crowdsourced reports from the community. In these cases, the researchers must adopt the platform-specific definition of hate speech.

## 2.2 Observable Factors of Hate Speech

Any operational approach that aims at using CS technology to recognize (automatically or semi-automatically) hate speech in digital communications must have procedures capable to systematically flag hate speech instances, somehow. Such procedures are expected to operate over public repositories, e.g., data bases of pieces of chats and comments that one can find appended to news, on-line games, and social media. They should be able to draw conclusion from what it is available in those repositories and from what it is observable in on-line chats and discourses. They cannot, instead, access to more intangible clues such as one's deliberation and/or premeditation in obtaining a certain effect. These latter ones, which are factors of relevance in other settings such as in a court trial, seems hardly observable on-line, although up to a certain point they may be deducible from a transcript set. Automatic or semi-automatic procedures are also unlikely to get direct access to elements that in psychology are relevant to the studying of hate such as, for instance, one's fear, obedience to authority, trust, sense of belonging (see Greisch, 2020; Sternberg, 2020); nor, we argue, it seems to be possible to observe variables such as belief, intentions, and desires which in formal logic and artificial intelligence are used to characterize and model social phenomena partially liked to hate, such as one's lying, bullshitting, persuading, deceiving others (Sakama, Caminada, & Herzig, 2014; Singh & Asher, 1993). Even Definition 1 cites factors which are not easily observable or reliably deducible from the sheer observation of digital repositories of speeches. For instance, it appears hard to find an evidence in the mere speech act that one's *attack* to one subject's or a subject group's "race, colour, religion, descent or national or ethnic origin" (as the EU Council Framework Decision 2008/913/JHA puts it) does in fact *instigate* hate and *incite* violence in others and against the same subject(s).

The difficulty of this task (i.e., to detect the real effect of one's act of speech over someone else) is probably one of the reasons why, despite the investments to mitigate the hate speech phenomenon, the attempts made by Twitter, Facebook, and YouTube are still criticised as unsatisfactory (Zhang & Luo, 2018). At the matter of fact, counter-acting hate still requires a direct or indirect intervention and moderation by human reviewers. Human moderation, in addition to be unsustainable, remains prone to error and it is largely vulnerable to subjectivity and ultimately not fully auditable; and, such, are the automatic methods that depend on it.

At the light of these considerations it should be clear that today's most common methods of hate speech detection have strong similarities with strategies used to classifying documents and processing written texts. *Textual data*, written speeches or transcripts, are thus the "primary" observable source from which to start any investigation about hate speech although of course other sources are possible.

The simplest solution is the sheer searching for elements of hate in a discourse (see Section 3.1). The same primary source (i.e., written texts) can however be pre-processed and annotated to enable further kind of analysis. Annotated data sets and gold standard repositories can be built and used e.g., in machine learning approaches (see Section 3.2). Here the recognition of hate emerges from the knowledge inferred from a large set of samples usually annotated by human users. Humans can also be involved in preparing the training data sets for machines and other knowledge-based approaches. Once set up with the right dataset, machine learning enables investigation over more abstract features than those possible from a more direct linguistic analysis of the texts. Sentiment analysis is one of them, as well as topic labelling, language detection, and intent detection. Here "intent" is the goal of a conversation not the intention/deliberation of the subject to wish to cause an effect, e.g., a specific tangible and immediate reaction in its audience of readers against a person or a category of people.

Whereas textual data about speeches are the most common primary raw data of an investigation, *videos and images* can be additional observable factors in the search for presence of hate on-line. To circumvent measures that look their presence in texts, certain hateful messages are hidden as visual text in images; as well, images or videos themselves can serve the purpose to instigate hate, but an analysis of images/videos in this sense seems at the moment be far-fetched for the current technology.

Another source of observable data are *metadata*. Metadata may not characterize directly the hateful nature of a content, but they can provide helpful information about its context. For instance, MacAvaney et al. (2019) report that "demographics of the posting user, location, timestamp, or even social engagement on the platform can all give further understanding of the post in different granularity" (p. 7). Metadata, we believe, can also help study how news spread across different media and social groups. They can be a valuable data set for the construction of models of social information flow and patterns, like the sort of epidemic models used in the analysis of malware. Hate news can follow a similar evolution.

## 2.3 Tasks Solved in CS

From an analysis of existing research works addressing hate speech from CS perspective, we have identified three main groups of approaches:

-   *Characterization*. In this class we have found studies on the understanding of hate speech. These works analyse characteristics related to hate speech content and authors who contribute such content. The existing works focused on influence of anonymity and differences across countries (Mondal, Silva, & Benevenuto, 2017), abusive users and their behaviour (Chatzakou et al., 2017) or dynamics and diffusion of hate speech (Mathew et al., 2019).

- *Detection*. This group embraces methods for detection of hate speech. The existing works focused on distinguishing hate speech from other types of offensive language (Davidson et al., 2017) or classification of hate targets (Zampieri et al., 2019; Liu et al., 2019). The approaches utilize many techniques from data science, especially machine learning.
- *Mitigation*. This class contains works on prevention and elimination mechanisms. Existing works focused on policies applied by social media platforms (Pater et al., 2016), on how such mechanisms are utilized in practice by adolescents (Wachs et al., 2020) or how the output from detection methods can be used to improve moderation of on-line discussions (Švec et al., 2018).

These three groups of approaches are closely connected: the results of characterization provide a valuable input for detection methods and consequently detection methods are crucial to automatically or semi-automatically support mitigation mechanisms.

So far, the CS research papers have been systematically reviewed in two surveys, i.e. Schmidt and Wiegand (2017) and Fortuna and Nunes (2018). In 2017, Schmidt and Wiegand analysed existing approaches for hate speech detection only. In 2018, Fortuna and Nunes provided a state-of-the-art overview of research papers addressing detection as well as characterization of hate speech. The previous surveys do not comment on mitigation.

In the next section, we focus specifically on hate speech detection, which is addressed by the majority of existing research works and which represents the most important task in addressing hate speech with data science.

## 3 Automatic Hate Speech Detection

Automatic hate speech detection is perhaps the most well researched task in CS. The goal of automatic hate speech detection is to create an algorithm that is able to process a content (the existing approaches focus on textual content, such as a social media post or a discussion comment) and make a prediction about whether it contains hate speech or not. In the simplest form this might be a simple binary *yes* or *no* decision (Davidson et al., 2017). Advanced versions might also be able to detect the target of the hate speech, i.e. particular individuals or groups (Zampieri et al., 2019; Liu et al., 2019).

There are two basic types of hate speech detection algorithms: the first type includes *lexicon-based algorithms*. They use the so-called hate speech lexicons. These are lists of words and phrases that are often found in hate speeches both general and aimed at specific target groups. This kind of hate speech detection is reduced to searching for lexicon entries or their derivatives. The second type, *machine learning*, can be used to create a hate speech detection model from a set of annotated texts, i.e. a collection of texts that are manually evaluated by humans. Machine learning algorithm can be used to process these data and create a model that is able to detect hate speech by applying the knowledge it learnt from the data using statistical methods.

Note the difference between how the hate speech is characterized between these two approaches. In the first case, we have a lexicon of words and phrases, in the second case we have a set of annotated examples.

We can also distinguish between approaches that only look at the text itself, i.e. the text is evaluated in isolation and those which take the metadata and context into account, e.g. who has written the text (Waseem & Hovy, 2016), how other users interacted with the text (Davidson et al., 2017), etc.

Regardless of the approach, any solution depends on a suitable dataset; in general, datasets represent a fundamental building block in data-science tasks. Automatic hate speech detection is not an exception.

### 3.1 Lexicon-based approaches

Hate speech lexicon is a set of words or phrases that are usually linked to hate speech. These lexicons are mostly hand-crafted, and they contain various racial, homophobic and religious slurs, and other types of slurs for different hate speech target groups. The task of hate speech detection is then reduced to searching for lexicon entries in a text.

Lexicon-based approaches may have a high precision (if lexicon is not too broad), i.e. most of the text that contains hate speech related slurs is indeed hate speech. Once we have a lexicon ready, it is easy to develop and deploy a hate speech detection method. Lexicon-based approaches are also easily interpretable, i.e. we can exactly tell why the method's internal mechanism decided to flag particular text as hate speech. *Interpretability* is an important property in the so-called *explainable AI*, i.e. AI methods that provide a human-understandable explanation of a decision taken automatically. Since on-line hate speech is, at least in EU, related to the wider category of hate crime and detection may be followed by punishment, explainability can become a measure to avoid abuses against freedom of speech and to rebut in case of errors.

On the other hand, lexicons are limited by our ability to maintain the ever-growing list of slurs used for various target groups. The limitation is even more severe in a multilingual setting, when we need to maintain such lexicons for multiple languages and cultures at the same time. Lexicon-based approaches are also context-insensitive. The presence of a slur does not necessarily mean hate speech: Davidson et al. (2017) showed that only a small percentage of samples marked by a lexicon were considered by human annotators as actual hate speech. Some slurs can be used in friendly manner, some slurs can mean something completely different in different languages or in different cultures. Lexicons also cannot detect hate speech that does not contain slurs at all. This causes low recall scores for lexicon-based approaches, i.e. many samples that contain hate speech are not flagged because they do not contain any of the slurs from the particular lexicon we use.

The most popular lexicon for hate speech detection is *Hatebase* (https://hatebase.org/). It is a continuously maintained database of hate speech terms for 95 languages. It has been already used in multiple works (Chatzakou et al., 2017; Mathew et al., 2019; Mondal et al., 2017). Other works create their own hate speech lexicons, sometimes in semi-automatic fashion (Gitari, Zuping, Damien, & Long, 2015).

### *3.2 Machine learning approaches*

Machine learning is an artificial intelligence field that deals with statistical models that are able to detect reoccurring patterns in data and to apply such patterns to predict some information for new data samples. The training of machine learning models is done with annotated data, i.e. data where we have a label assigned to each data sample. In the case of hate speech detection, we can have a collection of discussions posts, tweets or other type of textual/multimedia content that is annotated manually by human annotators. Such annotations say whether or not the sample contains hate speech and/or what kind of hate speech does it contain. The machine learning model processes data like these with statistical methods. The result is a parametric model (model whose behaviour is governed by the values of its parameters) that responds to the hate speech contained in the text.

### 3.2.1 Features

Most machine learning models expect a sample to be represented as numerical *features*. Feature engineering is a process of converting an original raw content (typically text of the sample but also its metadata and context) into features (i.e., numerical representation that can be processed by machine learning models). We can use various types of features for hate speech detection, for example features representing words used in the comment, sentiment of the comment or author's previous contributions.

The most basic features are *lexical features*. They encode what words are present in the text, e.g. we might have a binary vector where a zero or one at position $i$ says whether the $i^{th}$ word in a referenced wordlist is or is not present in the text. The model then learns whether some words or combination of words constitute an increased probability of hate speech. This approach can be extended by using various sequences of $n$ words (called n-grams), instead of single words. Term frequency-inverse document frequency (TF-IDF) is a popular kind of lexical representation, that also takes into account the frequency of word in a sample and the frequency of the word in the whole corpus to create a numerical representation that is more precise than a simple binary vector.

Hate speech is often associated with negative emotions, such as hate or fear. Various scores that calculate *sentiment or emotional range* of the text can also be utilized as features for hate speech detection. These scores are provided by other machine learning models that are trained on their sentiment-related tasks. Similarly, we can have scores that measure the quality of the text, e.g. the grade level or readability such as the Flesch-Kincaid readability index.

Finally, in recent years, pre-trained *distributed representations* are the go-to representations for text processing. They are latent feature vectors calculated for words or higher lexical units (sentences or paragraphs) by pre-trained natural language processing models. They are the so-called hidden representations that are calculated by neural networks for other tasks, such as language modelling or machine translation. These representations can be repurposed for any task, including hate speech detection. They encode various aspects of text as perceived by the original neural model.

Distributed representations are always used for neural models (Ousidhoum et al., 2019; Švec et al., 2018), but they can be also utilized for other model types. Compared to other representation types they provide the richest semantic understanding of the text, i.e. they contain a lot of useful information about the text. The classical lexical features are also being used (Chandra et al., 2017; Davidson et al., 2017; Ousidhoum et al., 2019; Zampieri et al., 2019) but compared to the distributed representation they achieve a lower performance. Finally, (Davidson et al., 2017) also use various sentiment-related and text quality related scores to enrich the lexical features.

### 3.2.2 Models

Model is an artefact created during machine learning that is able to perform a task, in our case hate speech detection. A hate speech detection model takes feature representation of a content as input, and it outputs a prediction about the hate speech contained within the sample. Models are trained iteratively by being exposed to annotated samples. Model tries to learn from these samples by updating its inner working so it can correctly predict whether the samples contain hate speech or not. Model trained this way can be consequently applied to previously unseen samples to obtain expected predictions.

*Linear models* are models that are trying to establish a so-called linear boundary between samples that contain and do not contain hate speech. This boundary is a hyperplane that lies in the feature space. Training of models like these is the process of moving this boundary in this space so that samples with hate speech lie on one side of this boundary while the samples without hate speech lie on the other side. *Support vector machines* are a popular extension that can work even with linearly inseparable feature spaces by adding additional dimensions.

*Decision trees* are machine learning models that learn to create a sequence of decisions that lead to a prediction. The decisions are iteratively formed and linked to create a logical sequence.

*Neural models* are the recently popular machine learning models based on an idea of abstracting the computational model of biological neurons. Multiple (up to millions) of artificial neurons form a so-called neural network – a trainable model that make predictions from features and learn from the data using gradient-based methods of optimization. They also support pre-training of parameters; a transfer learning is a technique where models trained on other tasks (most notably language modelling) are later repurposed for other tasks, such as hate speech detection. Neural models are notoriously uninterpretable, i.e. we cannot really say why the neural network decided to flag a text as hate speech. However, various approaches improving the interpretability were proposed already (Švec et al., 2018).

All the models we mentioned before are used for hate speech detection as well. Linear models (Davidson et al., 2017; Ousidhoum et al., 2019), support vector machines (Zampieri et al., 2019) and decision trees (Davidson et al., 2017; Liu et al., 2019) are the classical machine learning algorithms often used in natural language processing. However, recently neural networks were able to beat state-of-the-art results for many natural language processing tasks, including tasks related to hate speech detection, such as sentiment analysis, emotion analysis, etc. The performance improvement was also already shown in hate speech detection (Zampieri et al., 2019), even though we had not yet seen some of the state-of-the-art techniques, such as pre-trained language models, being applied.

### 3.3 Datasets

In general, CS researchers have two options how to obtain datasets, which are necessary to propose/evaluate their solutions (a) by utilization of existing datasets; or by (b) creating and labelling an own dataset. It is necessary to say, that there is no widely established benchmark dataset for hate speech detection task. Following the positive experiences from other data-science tasks, the presence of such benchmark dataset can have a positive influence on progress in the area as well as make performance comparison between individual solution easier and much representative.

### 3.3.1 Existing datasets

We are not aware of any hate speech datasets created and published (with a proper level of user anonymization) directly by owners of social media platforms. However, some existing datasets are published by CS researchers as the part of their papers, e.g., a dataset published by (Davidson et al., 2017); for an overview of additional datasets, please, see (Fortuna & Nunes, 2018). Many researchers, however, prefer not to disclose their datasets, and thus they remain unavailable for future research or for methods' performance comparison.

   Among data scientists, attending different types of competitions and challenges represents a popular form of research and innovation discovery. Some of them already focused on hate speech (and other types of offensive language) and thus they represent another source of existing datasets. Namely, they were conducted on Kaggle portal (https://www.kaggle.com/c/detecting-insults-in-social-commentary) as well as a part of SemEval tasks: SemEval-2019 Task 5 was focused on multi-lingual detection of hate speech against immigrants and women in Twitter (https://www.aclweb.org/anthology/S19-2007/), the latest SemEval-2020 Task 12 was focused on multilingual offensive language identification in social media (https://sites.google.com/site/offensevalsharedtask/). In all cases, human-labelled datasets were created. Utilization of such datasets brings a valuable possibility to compare the proposed method with the previous solutions submitted to the competition.

### 3.3.2 Collecting and labelling own datasets

The second option is to collect and label an own dataset. In this case, data science researchers must adhere to specific conditions under which the data are licensed and provided by social media platforms (typically by API). Most of social media platforms declare published content as their or user's property and thus it is not possible to do research on such data or publish them as datasets, without an explicit consent. On the other hand, some social media, for instance Twitter, allow collecting data for research purposes. For this reason, the most common source of own datasets for hate speech detection is Twitter (Chatzakou et al., 2017; Mondal et al., 2017; Zampieri et al., 2019). Nevertheless, researchers are still not allowed to publish large-scale datasets. Twitter's license conditions allow openly publish dataset for research purposes up to 10,000 samples. For larger datasets, only tweet identifiers may be published (without textual content), which could be used to obtain tweets also with additional complementary data.

Unfortunately, due to license issues other large social media platforms, such as YouTube or Facebook are used in research papers only very rarely (Fortuna & Nunes, 2018). Some researchers therefore decide to use the smaller social networks as a source for their own datasets, such as Whisper (Mondal et al., 2017), which provide its users significant anonymity or Gap (Mathew et al., 2019), which provide "free speech" also for users, who would be banned from other social media platforms due to violation of policies.

## 4 Challenges and limitations in computer science research

Addressing hate speech with data science is a complex and challenging task in many aspects. At the very beginning, since there is no single definition and categorization of hate speech in the CS literature, researchers as well as industry adopt different views. It would be helpful to have definitions of hate speech that leads to detection and mitigation methods equally applicable across the different social media. In order to make it consistent with other research domains (such as psychology or sociology), multidisciplinary contributions of various researchers may be highly valuable. Such unified widely adopted definition will result in many advantages, e.g. to make knowledge about hate speech as well as trained machine learning models more easily transferable between social media platforms.

Moreover, in the case of hate speech detection, we can observe specific challenges related to training machine learning models.

-   *Distinguishing between hate speech and other types of offensive language*. While distinguishing between benign language and hate speech achieves relatively a good performance, one of key challenges is a separation of hate speech from other types of offensive language (Davidson et al., 2017).

- *Dynamics of hate speech*. Hate speech is not static and it evolves quickly and heterogeneously in different languages and cultures. There is also a strong motivation of authors to find new ways how to hide hate speech and thus bypass existing human-based or (semi-)automatic detection approaches.
- *Spread of hate speech*. Studying how a particular piece of news or a speech spread through and across the various social media and from where it has its origin, may reveal specific patterns that can be later investigated, e.g., using model similar to those used to detect malware infections from their propagation patterns (e.g. del Rey, 2015). However, coming out with models that can reliably distinguish hate from other form of more non-aggressive discourses (e.g., gossips, and fake news) is at the moment future work. It may be unclear whether hate spreads differently than other forms of speech.

In addition, to address hate speech with data science, suitable datasets are crucial. Nevertheless, we can observe a persisting lack of such public and well-annotated data what results in many additional limitations and challenges:

- *Lack of public datasets*. Probably, the largest limitation of hate speech research is absence of large, context-rich and benchmark datasets, which will be public and freely available for all researchers. While only one existing research paper used an existing publicly available dataset, the majority of hate speech detection methods are trained and evaluated on own datasets (data from Twitter platform being used the most; Fortuna & Nunes, 2018). Unfortunately, only few of such own datasets are consequently made public. In addition, there is a limited research on data from some social media platforms (e.g., YouTube or Facebook). It is probable that some in-house algorithms are trained on such data, however, they are not covered by research publications.
- *Problems with copyrights*. One of reasons contributing to lack of public datasets is license issues related to usage of data from social media platforms. For example, Twitter conditions allow to publish only identifications of tweets without their particular content (authors working with such datasets are therefore required to use Twitter API to obtain tweets' content by themselves). However, since tweets with hate speech should be removed, many times such tweets are not available anymore and thus such datasets cannot be fully reconstructed.
- *Difficult annotation process*. Human-based manual annotation of hate speech is challenging mainly due to a low expert-agreement, a high time consumption and high demands on (cultural, sociological, etc.) expertise (Waseem, 2016). The low expert agreement is caused not only by unclear definitions of hate speech and arguable differences between hate speech text and non-hate speech text but also by the subjectivity of annotators (Waseem, 2016). Different people can view the same text from different perspectives, what can bring mislabelling of dataset samples. This creates a need for labelling datasets by multiple annotators and computing inter-agreement between them. While increasing number of annotators can help to reveal controversial samples from datasets, it also significantly increases resources and in final the labelling process is also much more expensive.

The mentioned limitations related to datasets significantly slow down a research progress and make comparison of individual approaches very problematic (Fortuna & Nunes, 2018). In order to reveal a high demand for datasets, we previously proposed and developed an infrastructure for monitoring antisocial behaviour on the web called Monant (Srba et al., 2019). The Monant infrastructure is continuously developed and currently it allows to obtain discussion posts from various news portals and discussion tools. Besides the content of discussion posts, also rich contextual pieces of information are extracted (e.g. news articles which are associated with a discussion).

## 5 Open Problems

Following the stated challenges and limitations in current CS research, we identified open problems and future directions as follows:

- *Multi-aspect hate speech detection*. The majority of existing approaches focus on yes/no (binary) detection of hate speech, that may not be sufficient in some cases (Ousidhoum et al., 2019). Recently, researchers started to distinguish various aspects of hate speech more precisely, such as an attribute which discriminates target group (e.g., religion, race, disability and sexual orientation: Liu et al., 2019; Ousidhoum et al., 2019) or target type as individuals or groups (Ousidhoum et al., 2019; Zampieri et al., 2019). Future work should continue in this trend and distinguish various aspects of hate speech and thus finally contribute to more effective analysis and mitigation of hate speech.
- *Multilingual and low-resource (minor) languages*. The research on multilingual hate speech is still in its beginnings (Ousidhoum et al., 2019). At the same time, multilingual hate speech detection provides an opportunity how to deal with lack of extensive datasets. For example, transfer learning provides an interesting research potential to take advantage of larger datasets (mostly in English language) while training models for minor languages, where potential datasets are very sparse and small.

- *Detection utilizing contextual information*. As soon as context-rich datasets will be available, a new research direction lies in research how feature engineering can be applied to derive innovative feature sets (e.g., social structures of authors contributing hate speech; Davidson et al., 2017) and how better models can be trained (e.g. with utilization of multi-view machine learning approaches).

- *Interpretable and explainable detection models*. Most times, the results of detection methods serve as an input for human moderators to check suspicious content and to take an action (e.g., to remove a content or to ban a user). Deciding whether to take such action may be difficult as some machine learning predictions (mostly those based on neural networks) are not easily interpretable. Research on methods how to provide interpretation of machine learning models (e.g., explaining decision by highlighting the part of content that contributed to the decision most; Švec et al., 2018) will be beneficial. Explainability is also a property that would allow to argue against any decisions taken against allegedly perpetrators of hate speech: automatic decisions can be biased too. In such cases, explainability would at least hold accountable AI and, in the long run, it may help also preserving freedom of expression by discouraging abuses of the instrument (e.g., for censorship).

- *Passive hate speech*. We have found no technical approach detecting hate when it is expressed in a passive aggressive style. Passive aggression, like sarcasms, are form of talking/writing which are not easily recognizable in written text, although specific figure of speech (e.g., hyperbole) or the typographic style (e.g., capitalized text) can reveal it up to a certain extent. We have found a work on detecting sarcasm that uses pre-trained models (Poria, Cambria, Hazarika, & Vij, 2017): the discerning is done with the support of humans. How to apply such techniques to reveal passive hate is an open problem.

- *Lack of mitigation mechanisms*. There are some research CS papers on characterization and detection of hate speech, but we found only very limited research how mitigation step should be performed. In this case, CS researchers may benefit from knowledge in related research domains (e.g., psychology or sociology) to propose new mechanisms how transfer results of detection methods to suitable prevention and mitigation actions.

- *Real-time hate speech checkers*. Eric Schmidt (2015), a former CEO of Google, suggested the use of a spell checker for harassment and hate language. In 2018, Facebook accidentally revealed a button for reporting the presence of hate speech (CBS News). It seems that those ideas remained in the drawer, never finding their way into an application; but having a "spell-checker" for hate could be helping users avoid at least some naive mistakes of "hate" style. Despite similar commercial tools for tone and mood (e.g., Grammarly; https://www.grammarly.com/), we could find nothing which is specialized for hate speech.

## 6 Conclusion

Hate speech without any doubts represents a significant problem that hinder effective and polite information and knowledge exchange in social media. Understanding, identification and prevention of hate speech is a substantial problem that is currently addressed by a multi-disciplinary research, in which computer science plays a significant role.

In this chapter, we focused on hate speech specifically from a computer science perspective and, in particular, from a data science viewpoint. We provided an overview how computer science contributes to three primary tasks: hate speech characterization, detection and mitigation. With this chapter, we aimed to present current state-of-the-art research in an easily understandable way, from which also researchers from other domains can benefit.

The analyses of computer science research papers (surveys as well as specific approaches) revealed the challenges behind addressing hate speech with information technologies, how the current approaches address such challenges and what are the open problems and the future directions.

## Acknowledgements
This work was partially supported by the Slovak Research and Development Agency under the contracts No. APVV-17-0267, APVV SK-IL-RD-18-0004; by the Scientific Grant Agency of the Slovak Republic, under the contracts No. VG 1/0725/19 and VG 1/0667/18; and by the FNR POC Project NoCry.